\documentclass[12pt]{iopart}
\usepackage{graphicx,color}
\usepackage{bm}
\usepackage[latin1]{inputenc}

\begin{document}

\title{Phase coherent transport in (Ga,Mn)As}

\author{D. Neumaier, K. Wagner, U. Wurstbauer, M. Reinwald, W. Wegscheider and D. Weiss}

\address{Institut f\"{u}r Experimentelle und Angewandte Physik,
Universit\"{a}t Regensburg, 93040 Regensburg, Germany}

\ead{daniel.neumaier@physik.uni-regensburg.de}

\date{\today}

\begin{abstract}

Quantum interference effects and resulting quantum corrections of
the conductivity have been intensively studied in disordered
conductors over the last decades. The knowledge of phase coherence
lengths and underlying dephasing mechanisms are crucial to
understand quantum corrections to the resistivity in the different
material systems. Due to the internal magnetic field and the
associated breaking of time-reversal symmetry quantum interference
effects in ferromagnetic materials have been scarcely explored.
Below we describe the investigation of phase coherent transport
phenomena in the newly discovered ferromagnetic semiconductor
(Ga,Mn)As. We explore universal conductance fluctuations in
mesoscopic (Ga,Mn)As wires and rings, the Aharonov-Bohm effect in
nanoscale rings and weak localization in arrays of wires, made of
the ferromagnetic semiconductor material. The experiments allow to
probe the phase coherence length $L_{\phi}$ and the spin flip length
$L_{SO}$ as well as the temperature dependence of dephasing.

\end{abstract}


\submitto{New Journal of Physics}

\maketitle

\tableofcontents

\section{Introduction}

The discovery of the ferromagnetic III-V semiconductor materials
(In,Mn)As \cite{Ohno1} and (Ga,Mn)As \cite{Ohno2} has generated a
lot of interest as these materials combine ferromagnetic properties,
typical for metals, with the versatility of semiconductors (for a
review see e.g. \cite{Ohno3,Ohno4,Ohno5,Dietl1,MacDonald}). This
allows, e.g., to control ferromagnetism by electric fields thus
opening new prospects for application and fundamental research
\cite{Ohno6}. The Mn atoms in the III-V host are not only
responsible for the ferromagnetism but also act as acceptors such
that, at sufficiently high Mn-concentration, (Ga,Mn)As is a
degenerate p-type semiconductor \cite{Oiwa}. The ferromagnetic order
of the Mn magnetic moments is mediated by holes via the
Ruderman-Kittel-Kasuya-Yosida (RKKY) interaction \cite{Dietl2}. By
now ferromagnetism in (Ga,Mn)As is well understood, allowing to
predict Curie temperatures \cite{Dietl3}, magnetocrystalline
anisotropies \cite{Sawicki} as well as the anisotropic
magnetoresistance effect \cite{Baxter}. In this respect (Ga,Mn)As is
one of the best understood ferromagnetic materials at all
\cite{Jungwirth} and hence suitable as a model system to study
quantum corrections to the conductivity in ferromagnets. The quest
to increase the Curie temperature $T_C$ in (Ga,Mn)As towards room
temperature has led to a thorough investigation of the material
properties (see, e.g. \cite{Jungwirth2} and references therein). By
annealing (Ga,Mn)As sheets or by incorporating them into
sophisticated layered arrangements the Curie temperature was
increased up to 173 K \cite{Edmonds,Wang} and 250 K with Mn-doping
\cite{Nazmul}, respectively. Despite the high crystalline quality of
the material (Ga,Mn)As is a quite disordered conductor on the verge
of the metal-insulator transition (MIT). For Mn concentrations on
the metallic side of the MIT the typical mean free path $l$ of the
holes is a few lattice constants. Hence it was until recently an
open issue whether quantum effects like Aharonov-Bohm (AB)
oscillations, universal conductance fluctuations (UCF) or weak
localization (WL) can be observed in this class of materials.
Accordingly the phase coherence length $L_\phi$ and the
corresponding dephasing mechanisms which govern quantum mechanical
interference phenomena in ferromagnetic semiconductors were unknown.
Apart from being a fundamental material parameter this information
is needed to design more sophisticated layered structures from
ferromagnetic semiconductors. Examples are resonant tunneling diodes
or other interference devices which rely on the electrons' wave
nature. Apart from the question on the relevant phase coherence
lengths, quantum corrections to the resistance like weak
localization are suppressed by a sufficiently strong perpendicular
magnetic field $B$ \cite{Bergmann}. Hence the question arises
whether such effects can be observed at all in ferromagnets which
have an intrinsic magnetic induction. Therefore, the advent of the
new ferromagnetic semiconductor material (Ga,Mn)As with
significantly smaller internal field compared to conventional
ferromagnets offered a new opportunity to address such questions.

Below we first describe experiments to explore universal conductance
fluctuations in (Ga,Mn)As nanowires to get knowledge of the relevant
dephasing length and the main phase breaking mechanism. We show
below periodic conductance oscillations in (Ga,Mn)As nanoscale rings
to prove phase coherent transport by the Aharonov-Bohm effect. The
last part of this manuscript describes weak localization experiments
in arrays of (Ga,Mn)As nanowires, where aperiodic conductance
fluctuations are suppressed by ensemble averaging.

\section{Sample preparation and measurement technique}

\begin{table}
\scriptsize \caption{Parameters of individual wires (w), rings (r)
and wire arrays (a). Length \emph{L}, width \emph{w}, thickness
\emph{t} and number of lines parallel \emph{N} of the wire samples.
The diameter $d$ gives the inner and outer diameter of the rings.
Some of the samples were annealed for the time \emph{a} at 200 °C.
Curie temperature $T_C$ and the carrier concentration \emph{p} were
taken from corresponding reference samples. The phase coherence
length of the samples w1-w5, r1 and r1a were calculated using
equation (1) with $C=0.41$, as we measure at high magnetic fields
\cite{Lee} and have spin-orbit interaction \cite{Chandrasekhar}
explained in the text below. The phase coherence length of sample r2
and r3 were calculated using the amplitude of the AB oscillations
and equation (2). For the wire arrays a1-a2 we fitted the weak
antilocalization correction given by equation (3). The phase
coherence length and the corresponding dephasing time were taken at
20 mK for all samples.}
\begin{tabular}{lllllllllllll}
\br
Sample&w1&w2&w3&w4&w5&r1&r1a&r2&r3&a1&a1a&a2\\
\mr
$L$ (nm)&100&200&300&800&370&-&-&-&-&7500&7500&7500\\
$w$ (nm)&20&20&20&20&35&30&30&18&30&42&42&35\\
$t$ (nm)&50&50&50&50&42&50&50&42&42&42&42&42\\
$N$&1&1&1&1&1&-&-&-&-&25&25&12\\
$d$ (nm)&-&-&-&-&-&120-180&120-180&120-155&160-220&-&-&-\\
$a$ (h)&0&0&0&0&0&0&12&51&51&0&51&0\\
$T_C$ (K)&55&55&55&55&90&?&100&150&150&90&150&90\\
$p$ ($10^{-5}\Omega$m)&1.8&1.8&1.8&1.8&3.5&2.7&3.1&9.3&9.3&3.8&9.3&3.8\\
$D$ ($10^{-5}$m$^2$/s)&4.8&4.8&4.8&4.8&8&2.5&4.4&12&12&8&12&8\\
$L_\phi$ (nm)&110&95&90&155&135&125&160&130&130&150&190&160\\
$\tau_\phi$ (ps)&270&195&170&500&175&625&600&140&140&280&300&320\\
\br
\end{tabular}
\normalsize
\end{table}

To explore phase coherent phenomena in (Ga,Mn)As we fabricated
nanoscale wires, rings and arrays of wires, connected in parallel,
from several wafers, always containing a ferromagnetic (Ga,Mn)As
layer on top of a semi-insulating GaAs (100) substrate
\cite{Reinwald}. The magnetic easy axis of all investigated wafers
was in plane. The relevant parameters of the samples used are listed
in table 1. All investigated samples were fabricated using
electron-beam-lithography (a Zeiss scanning electron microscope
controlled by a nanonic pattern generator) and dry etching
techniques. Ohmic contacts to the samples were made by thermal
evaporation of Au and lift-off after brief in-situ ion beam etching
for removing the native oxide layer of (Ga,Mn)As. The measurements
were performed in a toploading dilution refrigerator (a Oxford
Kelvinox TLM), equipped with a 19 T magnet, by standard four probe
lock-in technique. The temperature regime accessible with this
cryostat system ranges from 1 K to 15 mK. To avoid the heating of
charge carriers at very low temperatures careful wiring, shielding
and the use of low excitation currents (down to 10 pA) were crucial.
Without such measures  quantum interference effects in (Ga,Mn)As can
not be observed. The schematic experimental set-up is shown in
figure 1.  As we observe no saturation of the signals for the
different effects (UCF, AB-oscillations and WL) at the lowest
temperatures we assume that the effective electron temperature is in
equilibrium with the bath temperature even at temperatures as low as
20 mK.

\begin{figure}
\begin{indented}
\item[]\includegraphics[width=\linewidth]{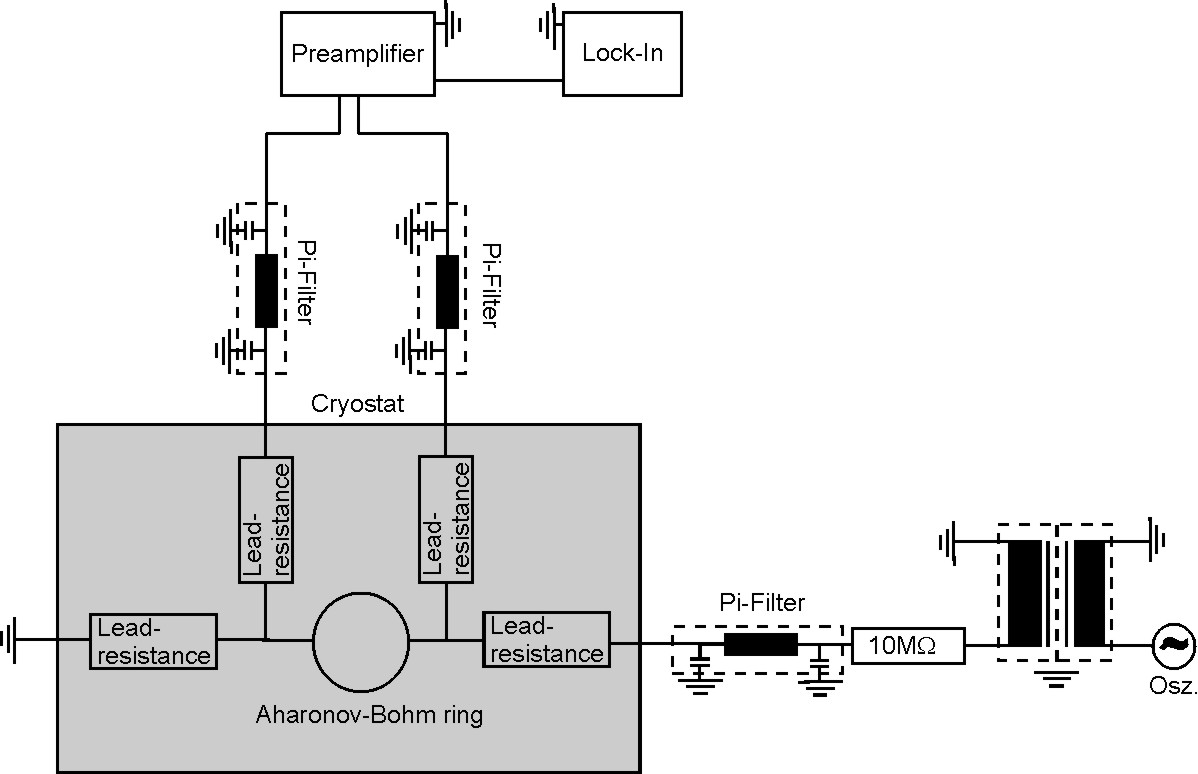}\\
\caption{Circuit diagram of our experimental setup (here with an
Aharonov-Bohm ring) used in the experiments.}
\end{indented}
\end{figure}

In the experiments described below the wires can be treated as quasi
one dimensional in the context of phase coherent transport. The
relevant parameters for the one dimensional treatment are the phase
breaking length $L_\phi$ and the thermal diffusion length $L_T$. The
phase breaking length gives the length scale an electron can travel
without loosing phase information. The thermal diffusion length
gives the length scale on which the thermal broadening of energies
around the Fermi surface leads to a smearing of the interference
effects. The thermal length is given by $L_T=\sqrt{\hbar D/k_BT}$.
Here $D$ is the diffusion constant given by
$D=\frac{1}{3}v_F^2\tau_P$, with the Fermi velocity $v_F$ and the
momentum relaxation time $\tau_P$. So our samples can be treated as
quasi one dimensional as long as their width $w$ (in our experiments
ranging from 18 nm to 42 nm) and thickness $t$ (ranging from 42 nm
to 50 nm) are smaller than $L_\phi$ and $L_T$. The thermal diffusion
length of our samples is $\sim 200$ nm at 20 mK. Below we show that
the dephasing length in our samples ranges between 100 nm and 200 nm
at 20 mK.

As we deal here with ferromagnetic material the actual value of the
magnetic field inside the material requires some attention. In the
experiments described below the external magnetic field is always
aligned along the growth direction of the (Ga,Mn)As structures,
denoted as z-direction. Taking the external magnetic field $B_z$
along this z-direction, the magnetic field inside the (Ga,Mn)As is
given by $B^\prime_z=B_z+J_z(1-N_z)$ where $J_z$ is the magnetic
polarization in z-direction and $N_z$ is the corresponding
demagnetization factor. In case of a two dimensional ferromagnetic
film, $N_z=1$ and internal and external field (in z-direction) are
identical. In case of wires, explored here, the demagnetization
factors are approximated by using cigar shaped ellipsoids with their
long axis along the wire axis and the short axes corresponding to
wire width/thickness \cite{Hubert}. For aspect ratios relevant here,
i.e., a wire length of order 1 $\mu$m (in x-direction) and wire
cross sections of about 40 nm we obtain for the demagnetization
factors $N_z=N_y\approx 0.5$ and $N_x=0$. Hence the internal field
is $B^\prime_z=B_z+J_z/2$. The maximum value of $J^{sat}_z$ is given
by the saturation magnetization of (Ga,Mn)As. Maximum values of our
samples' saturation magnetization are $\sim$30 emu/cm$^3$ which
translates in SI-system into values of $\sim$40 mT (see e.g.
\cite{Ohno5}). This means that the maximum difference between the
externally applied field $B_z$ and the field $B^\prime_z$ inside the
material is about 20 mT. The difference between $B^\prime_z$ and
$B_z$ is practically zero at $B_z=0$ and at most $\sim$20 mT at the
saturation field. This is only a small correction of the external
field in the sense that the difference between internal and external
field would hardly be noticeable on the magnetic field axes
displayed below. We hence use in the following that
$B^\prime_z\approx B_z=B$.

\section{Universal conductance fluctuations in (Ga,Mn)As wires and
rings}

One of the most straight forward methods of measuring the phase
coherence length and its temperature dependency relies on
measurements of universal conductance fluctuations in a single
mesoscopic wire. Universal conductance fluctuations stem from
interference of partial electron (hole) waves, scattered in a
disordered mesoscopic conductor (\cite{Washburn} and references
therein). If the wire is smaller than $L_{\phi}$ in all three
spatial dimensions the fluctuation amplitude $\delta G=\sqrt{\langle
(G-\langle G\rangle)^{2}\rangle}\approx e^{2}/h$, where the bracket
$\langle...\rangle$ denotes averaging over $B$ \cite{Lee}. The
amplitude of these fluctuations gets, in contrast to AB
oscillations, not exponentially damped once the wire length exceeds
the dephasing length, but attenuated by a power law \cite{Lee}:
\begin{equation}
\delta G_{rms}=C\frac{e^2}{h}\left(\frac{L_{\phi}}{L}\right)^{3/2}.
\end{equation}
\begin{figure}
\begin{indented}
\item[]\includegraphics[width=\linewidth]{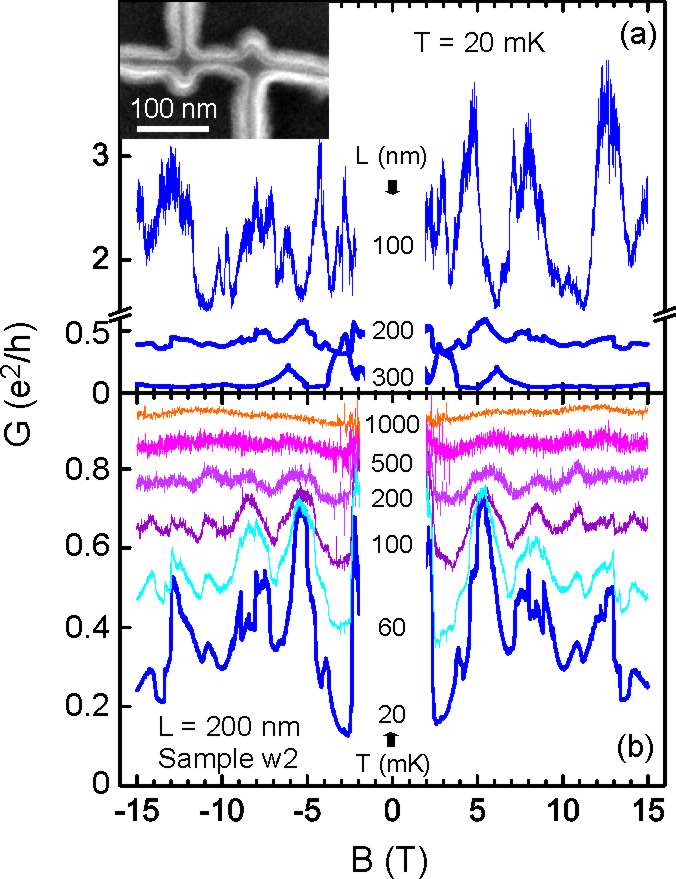}\\
\caption{(a) Magnetoconductance of three wires (sample w1, w2 and w3
in table 1) having different lengths $L$. The inset shows an
electron micrograph of sample w1 having a length of 100 nm and a
width of 20 nm. (b) Magnetoconductance of the 200 nm long wire w2
measured at different temperatures.}
\end{indented}
\end{figure}
Here, $L$ is the wire length and $C$ is a constant, with a value
close to or smaller than unity, depending, e.g., on the strength of
spin-orbit coupling \cite{Chandrasekhar} and the applied magnetic
field \cite{Lee}. Equation (1), describing the fluctuation amplitude
of one-dimensional (1D) conductors is applicable to extract the
phase coherence length as long as $L_\phi$ is larger than the width
$w$ and the thickness $t$ of the wire, and if the thermal diffusion
length $L_T$ is larger than $L_\phi$ \cite{Lee}: $w,t<L_\phi<L_T$.
Also the temperature dependency of the dephasing length can be
extracted by measuring the temperature dependence of the fluctuation
amplitude as $L_\phi$ is $\propto\delta G^{2/3}$. To investigate
UCFs in (Ga,Mn)As we fabricated individual nanowires with lengths
between 100 nm and 800 nm and measured their resistance $R$ in a
perpendicularly applied magnetic field $B$. The relevant parameters
of the wires, labeled w1-w4, are given in table 1. The conductance
$G$ was obtained by inverting the resistance: $G=1/R$. In figure 2a
the conductance of three wires having a length of 100 nm, 200 nm and
300 nm, respectively, is plotted as a function of the
perpendicularly applied magnetic field at 20 mK. Here, we focus on
the high field region ($B>2$ T) where the magnetization is saturated
and aligned along the external magnetic field. For all three wires
aperiodic, well reproducible conductance fluctuations are visible
\cite{Konni}. With increasing wire length not only the conductance
of the wires decreases but also the amplitude of the conductance
fluctuations drops, showing that at least the 200 nm and 300 nm long
wires are longer than the dephasing length $L_\phi$. The fluctuation
amplitude measured in the 100 nm wire is $\sim$0.5 e$^2$/h.
Depending on the exact value of the prefactor $C$ in equation (1)
the dephasing length is close to the wire length of 100 nm at 20 mK.
To investigate the temperature dependency of  the dephasing length
we measured the conductance fluctuations of the wires at different
temperatures. The magnetoconductance trace of wire w2 is displayed
in figure 2b for different temperatures between 20 mK and 1 K. With
increasing temperature the amplitude of the conductance fluctuations
is decreasing until they disappear above $\sim$200 mK. Plotting the
amplitude of the conductance fluctuations versus temperature in a
log-log-plot (figure 3) gives a power law for the temperature
dependence of the conductance fluctuations:  for the 200 nm long
wire w2 we obtain $\delta G\propto T^{-0.77}$. This temperature
dependency is similar for all investigated wires with an exponent
between -0.77 and -0.81, approximated in the following by -3/4.
Assuming a temperature independent prefactor $C$ we arrive at the
temperature dependency of the dephasing length: $L_\phi\propto
1/\sqrt{T}$. In case $L_T$ is smaller than $L_\phi$ the same
temperature dependency of $L_\phi$ results. Instead of equation (1),
$\delta G=C\frac{e^2}{h}\frac{L_T}{L}(\frac{L_\phi}{L})^{1/2}$ with
$L_T=\sqrt{\hbar D/k_BT}$ has to be used \cite{Lee}. Doing so we
again arrive at $L_\phi\propto 1/\sqrt{T}$. The inset of figure 3
shows the amplitude of the conductance fluctuations normalized by
the wire length. As all investigated samples lie on one straight
line the expected length scaling of UCF, given by equation (1), is
experimentally confirmed.
\begin{figure}
\begin{indented}
\item[]\includegraphics[width=\linewidth]{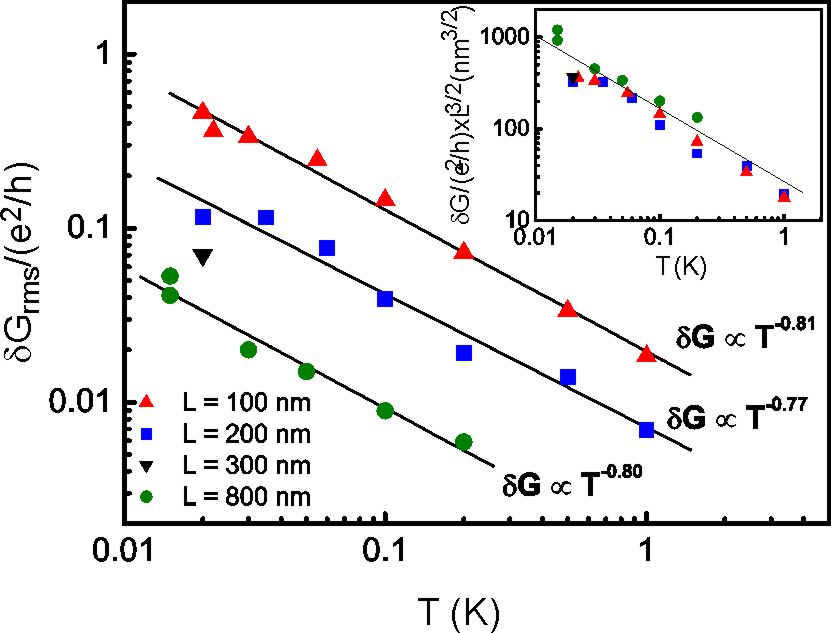}\\
\caption{Amplitude of the conductance fluctuations for the samples
w1, w2, w3 and w4 (see table 1) plotted in a log-log diagram. The
slope gives the temperature dependency of the conductance
fluctuations: $\delta G\propto 1/T^{3/4}$. The inset shows the
amplitude of the conductance fluctuations normalized by the wire
length.}
\end{indented}
\end{figure}

Before discussing this temperature dependence in more detail we
address the conductance fluctuations in the low field regime,
excluded in the discussion of figure 2. Corresponding data, taken
from sample w5 are shown in figure 4. The magnetic field scale on
which the conductance fluctuates is much shorter in the low-$B$
regime. The grey shaded area in figure 4 corresponds to the magnetic
field range where the anisotropic magnetoresistance effect is
observed, i.e., the $B$-range where the magnetization is rotated
from an in-plane to an out-of-plane orientation by the externally
applied magnetic field. The correlation field of the conductance
fluctuations is strikingly different in the low and high field
regime. Similar behavior was observed in previous experiments on
samples with in-plane easy axis \cite{Konni,Vila} and ad hoc
ascribed to the formation of domain walls \cite{Vila}. As in the low
field regime the magnetic configuration is continuously changed by
the external field, the scattering configuration is altered and the
correlation field, which can also be used to extract the phase
coherence length \cite{Lee,Konni}, is not any more a well defined
quantity. The phase coherence length extracted from the amplitude of
the conductance fluctuations is, within experimental error, the same
for low and high-field fluctuations.

The $L_\phi\propto 1/\sqrt{T}$ temperature dependency of the
dephasing length has been confirmed by others, also investigating
universal conductance fluctuations in (Ga,Mn)As nanowires
\cite{Vila}. Such a relatively weak temperature dependency for the
dephasing length is typical for electron-electron scattering as
dominating source of dephasing. In case of electron-phonon
scattering or electron-magnon scattering one would expect a stronger
temperature dependency of $L_\phi$ \cite{Sergeev,Takane}:
$L_\phi\propto 1/T^1...1/T^2$. At low temperatures and reduced
dimension electron-electron scattering with small energy transfer,
the so called Nyquist scattering, becomes more effective, leading to
$L_\phi\propto 1/T^{1/3}$ \cite{Altshuler}. However this temperature
dependency doesn't describe our results correctly. A possible
candidate for dephasing in (Ga,Mn)As nanowires might be critical
electron-electron scattering, describing dephasing in a highly
disordered metal near the metal insulator transition \cite{Dai}. The
corresponding phase coherence length depends like $L_\phi\propto
1/T^{1/2}$ on temperature. This is in accord with our results, but
the calculations were done for a 3D system. Hence the detailed
microscopic origin of dephasing in (Ga,Mn)As wires is still an open
issue.

\begin{figure}
\begin{indented}
\item[]\includegraphics[width=\linewidth]{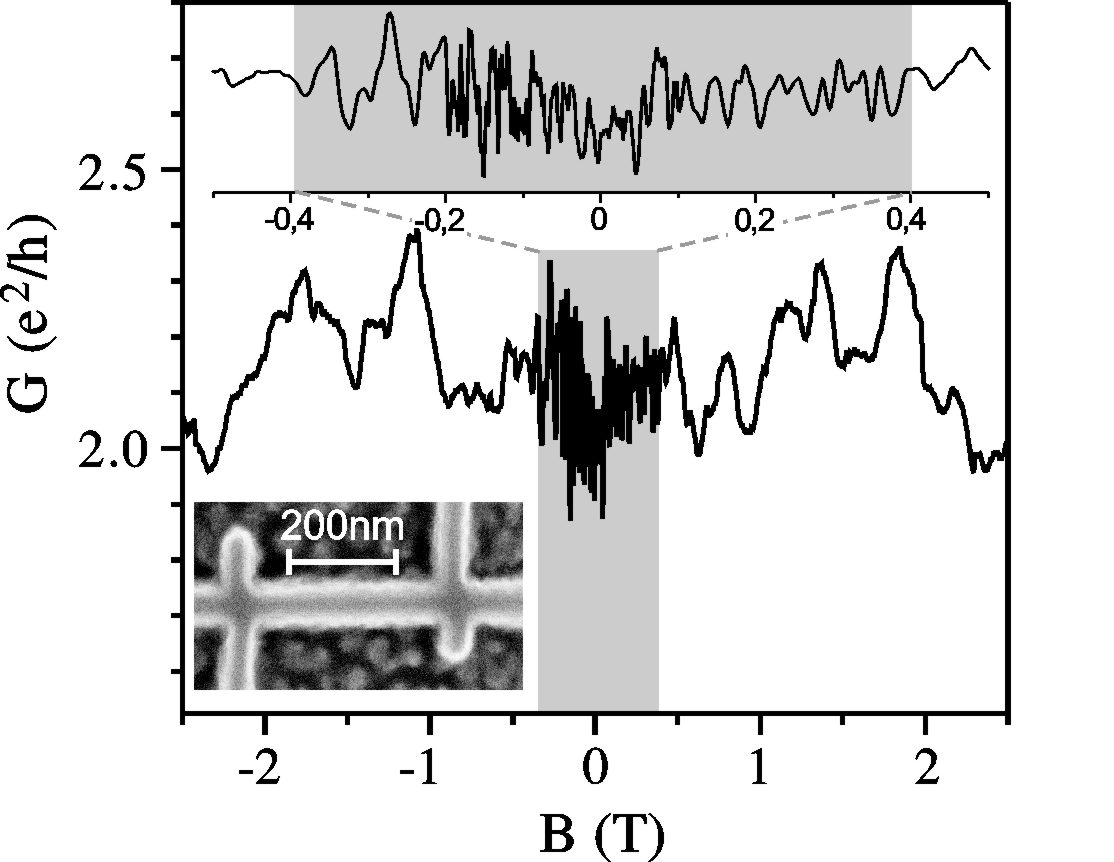}\\
\caption{UCFs measured in sample w5 (see table 1). An electron
micrograph of sample w5 is shown in the lower left inset. The grey
shaded regime corresponds to the magnetic field range where the
magnetization changes direction. The upper inset shows the low field
UCFs in an expanded magnetic field scale.}
\end{indented}
\end{figure}

To probe whether electron-electron scattering is the main source of
dephasing we investigated universal conductance fluctuations under
non equilibrium conditions, meaning that the effective electron
temperature is higher than the lattice temperature. To control the
effective electron temperature the applied excitation current has
been varied. Below we assume that even for our highest excitation
currents of 8 nA the lattice temperature is not increased. This
current corresponds to a voltage drop of 67 $\mu$V across the sample
and a heating power of $5*10^{-13}$ W. The cooling power of our
dilution refrigerator at 40 mK, on the other hand, is approx. 50
$\mu$W which is 8 orders of magnitude higher.
\begin{figure}
\begin{indented}
\item[]\includegraphics[width=\linewidth]{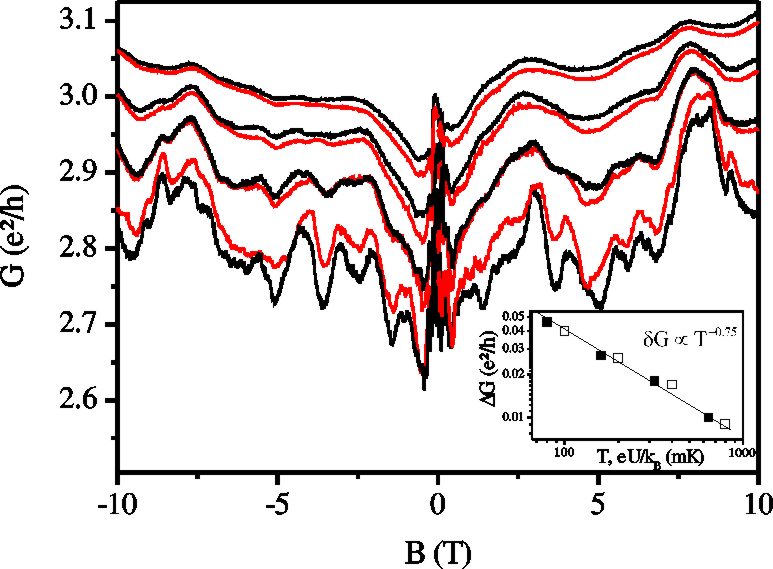}\\
\caption{Magnetoconductance of sample r1a measured at different
temperatures (black lines from top to bottom: 640 mK, 320 mK, 160 mK
and 80 mK) and at a fixed temperature of 40 mK at different
excitation currents (red lines from top to bottom: 8 nA, 4 nA, 2 nA
and 1 nA). The inset shows the amplitude of the conductance
fluctuations of sample r1a measured at different temperatures (solid
squares) and at a fixed temperature of 40 mK at different excitation
currents (open squares). In both cases we achieve the same
temperature dependency of the fluctuation amplitude: $\delta
G\propto 1/T^{3/4}$.}
\end{indented}
\end{figure}
The sample is located in the $^3$He-$^4$He mixture, so that thermal
coupling should be sufficient to keep the lattice in equilibrium
with the bath temperature. Figure 5  displays universal conductance
fluctuations in a ring with a diameter of 150 nm (sample r1a in
table 1). The conductance of this ring didn't exhibit periodic
conductance oscillations due to the Aharonov-Bohm effect, but only
aperiodic conductance fluctuations, similar to the ones observed in
wires. In figure 5 the fluctuations are successively suppressed by
increasing either temperature or current. In one experiment the bath
temperature was varied from 80 mK to 640 mK (black lines from bottom
to top), while the excitation current was kept at a value where the
effective electron temperature is still in equilibrium with the bath
temperature (200 pA at 80 mK, 500 pA at 160 mK and 320 mK, and 1 nA
at 640 mK. For comparison: 1 nA corresponds here to a voltage drop
$U$ of 9.5 $\mu$V across the sample, equivalent to $eU/k_B=110$ mK).
In this regime, the conductance fluctuations do not depend on the
excitation current as long as the hole gas and the lattice are in
equilibrium. Then we varied the excitation current, at a fixed bath
temperature of 40 mK, from 1 nA to 8 nA (red lines from bottom to
top). In that case the fluctuation amplitude is depending on the
excitation current. As one can see in figure 5 the black and red
traces lie quite well on top of each other. Thus we arrive at the
same experimental result in two ways, first by increasing the bath
temperature and secondly by raising the excitation current. Since
the excitation current can't change the lattice temperature, as
argued above, the observed dephasing can't depend on the lattice
temperature. This means that electron-phonon and electron-magnon
scattering can be excluded as source of dephasing. This is also
consistent with the temperature dependency of the dephasing length,
discussed above. The inset of figure 5 shows the average amplitude
of the conductance fluctuations in figure 5, both as a function of
the bath temperature (solid symbols) and the applied voltage in
units of $eU/k_B$ (open symbols), respectively. Both traces display
the same slope and lie above each other. This suggest a linear
correlation of effective temperature and applied voltage across.
Such a linear dependence is only expected if the lattice is
decoupled from the electrons \cite{Wellstood} and if the effective
electron temperature is given by $eU/k_B$. The dominating parameter
for dephasing in the low temperature regime is hence the effective
electron temperature and electron-electron scattering is the most
likely source of dephasing.

Low temperature annealing of (Ga,Mn)As causes an out diffusion of Mn
ions which occupy interstitial lattice sites, act as double donors
and couple antiferromagnetically to Mn ions on regular Ga-sites. So
low temperature annealing increases the carrier concentration, the
Curie temperature and the total magnetic moment of the samples
\cite{Edmonds}. Has such low temperature annealing of the samples an
influence on the phase coherence length? In nonmagnetic metals
magnetic impurities strongly reduce the phase coherence length
\cite{Pierre}. This poses the question whether Mn interstitials have
a similar effect on the dephasing length in (Ga,Mn)As, as they
increase the magnetic disorder due to their random distribution. In
order to look for effects of low temperature annealing we measured
the conductance of a 150 nm diameter ring at 20 mK before (r1) and
after annealing (r1a). The corresponding magnetoconductance traces
are shown in figure 6 before (black line) and after annealing (red
line). As expected the conductance increased significantly after
annealing (here by a factor of approx. 2). While the amplitude of
the conductance fluctuations is 0.055 e$^2$/h before and 0.080
e$^2$/h after annealing, the corresponding dephasing length,
determined by equation (1), is increased by 30 \% after annealing.
Calculating the dephasing time $\tau_\phi=L_\phi^2/D$ we find that
the dephasing time does not change after annealing. While $L_\phi$
increases, also the diffusion constant $D=\frac{1}{3}v_F^2\tau_P$
increases by about 75 \% since both, the Fermi velocity $v_F$
(increased hole density) and the momentum relaxation time $\tau_P$
get larger. This means that the dephasing time is essentially not
affected by low temperature annealing. The change in coherence
length is only due to the change of the diffusion constant. If
electron-electron interaction is the source of dephasing one would
expect that $\tau_\phi$ depends on the carrier concentration. Since
in our experiment the change in carrier density was only 15 \% the
$\tau_\phi (p)$ dependency could not be resolved. The main
difference of the Mn interstitials in (Ga,Mn)As compared to magnetic
impurities in normal metal is the coupling strength. In normal
metals the magnetic impurities are uncoupled above the
Kondo-temperature and a spin-flip process is energetically
accessible for the electrons, leading to a loss of phase information
\cite{Webb}. In (Ga,Mn)As the Mn interstitials are coupled quite
strongly \cite{Jungwirth} and a spin-flip process is not possible.
This explains why low temperature annealing does not increase the
dephasing time, but only increases the dephasing length by an
increase of the diffusion constant.

\begin{figure}
\begin{indented}
\item[]\includegraphics[width=\linewidth]{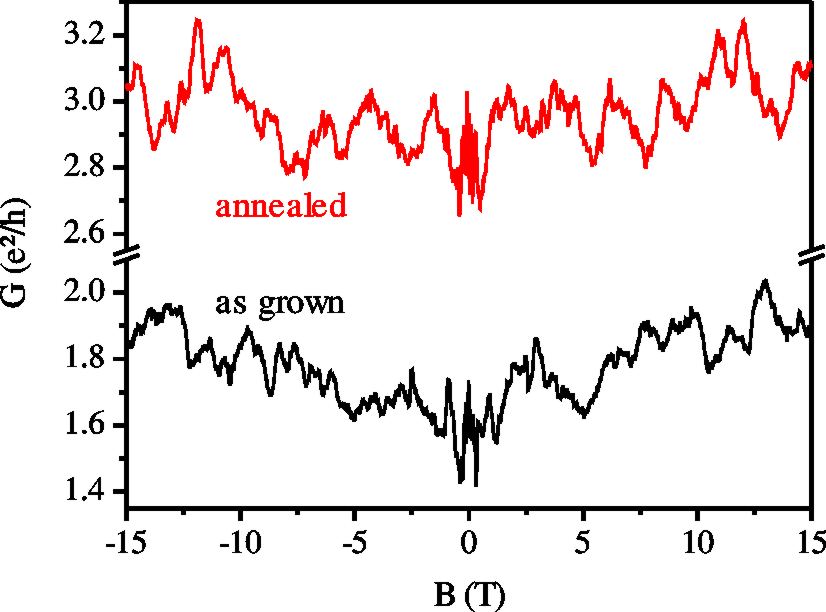}\\
\caption{Magnetoconductance of a ring before (sample r1) and after
annealing (sample r1a) measured at 20 mK.}
\end{indented}
\end{figure}

Spin-orbit (SO) interaction plays an important role in the valence
band of (Ga,Mn)As and hence also has an effect on the analysis of
UCFs. In equation (1) SO-interaction was not taken into account.
Chandrasekhar et al. pointed out that the presence of spin-dependent
scattering affects the amplitude of the conductance fluctuations
\cite{Chandrasekhar}. The amplitude of the conductance fluctuations,
determined by equation (1), depends on the ratio of phase coherence
length $L_\phi$ and spin orbit lengths $L_{SO}$. With increasing
$L_\phi/L_{SO}$ the amplitude of the UCFs gets reduced. This means
that in the presence of spin orbit interaction the phase coherence
length gets larger for a given value of the UCF amplitude. To
determine the corresponding factor requires knowledge of the
spin-orbit length $L_{SO}$ which can be extracted from
weak-antilocalization correction. After we have extracted $L_{SO}$
from weak localization experiments, discussed below, we return to
this point again.

\section{Aharanov-Bohm effect in (Ga,Mn)As rings}

\begin{figure}
\begin{indented}
\item[]\includegraphics[width=\linewidth]{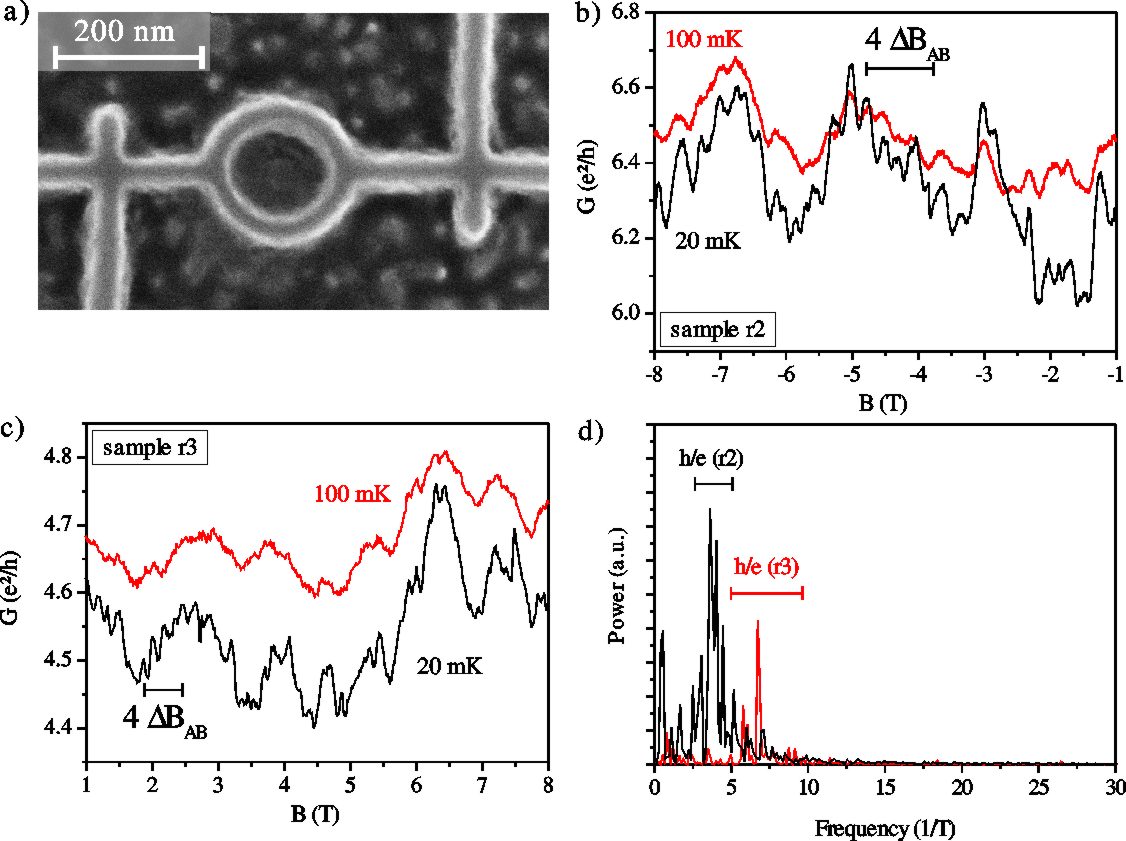}\\
\caption{a) Electron micrograph of sample r3 having an average
diameter of 190 nm. b) Magnetoconductance of sample r2 in a
perpendicular applied magnetic field measured at 20 mK (black line)
and 100 mK (red line). The period corresponding to the Aharonov-Bohm
effect is marked. c) Magnetoconductance of sample r3 in a
perpendicular applied magnetic field measured at 20 mK (black line)
and 100 mK (red line). The period corresponding to the Aharonov-Bohm
effect is marked. d) Fourier transformation of the
magnetoconductance at 20 mK of sample r2 (black line) and sample r3
(red line). The magnetic field interval expected for h/e
oscillations, evaluated by the inner and the outer diameter of the
ring, is marked.}
\end{indented}
\end{figure}

A quite prominent manifestation of phase coherence in mesoscopic
conductors is the so called Aharonov-Bohm (AB) effect. In a ring
geometry a wave-packet is split into two partial waves propagating
along the upper and lower half of the ring perimeter and interfering
at the "exit". The phase of the partial waves can be tuned by a
perpendicular magnetic field so that, as a function of magnetic
field strength, constructive and destructive interference of the
partial waves can be adjusted. At sufficiently low temperatures the
conductance across the ring displays periodic oscillations with a
period $\Delta B=\Phi_o/A$, were $A$ is the area enclosed by the
ring and $\Phi_0$ is the flux quantum h/e. Periodic Aharonov-Bohm
oscillations have been observed over the last decades in normal
metals \cite{Washburn2} and in low dimensional electron and hole
systems, e.g. \cite{Wegscheider2,Garbic}. However, the observation
of AB-oscillations in ferromagnetic ring-structures is even more
challenging and until quite recently it was unclear whether
AB-effects can be observed at all in ferromagnets. First observation
of AB effects have been reported in FeNi rings in the year 2002
\cite{Kasai}. Measuring the AB effect in (Ga,Mn)As-rings requires
that the phase of a wave packet is preserved while traversing the
ring. With increasing ring radius $r$ the amplitude is damped
exponentially, $\delta G\propto $exp$(-\pi r/L_\phi)$
\cite{Washburn2}. Additionally an aspect ratio close to one (ratio
of inner and outer radius of the ring) is needed to avoid smearing
of the interference pattern. In (Ga,Mn)As the phase coherence length
is about 100 nm at 20 mK. Hence AB oscillation should be visible in
rings with diameters of about 100 nm to 200 nm. The conductances of
the rings r1 and r1a, discussed above in figure 5 and 6 display no
clear AB oscillations. In an attempt to resolve AB-type of
oscillations the ring diameter was further reduced to a diameter of
only 100 nm \cite{Konni}. Corresponding magnetoconductance data of
this ring clearly containing a contribution of the AB effect, are
published in Ref. \cite{Konni}. However, the associated Fourier
spectrum is less clear as it displays no clear peak. One reason for
this is the relatively bad aspect ratio of the ring causing a
broadening of the Fourier peak. In an attemp to resolve AB
oscillations we tried to increase the phase coherence length by
increasing the diffusion constant of the (Ga,Mn)As material. We
found the maximum diffusion constant of $12*10^{-5}$ m$^2$/Vs in an
optimally annealed wafer containing approx. 5-6 \% manganese and
having a Curie temperature of 150 K. From this wafer we fabricated a
ring with an inner diameter of 120 nm and an outer diameter of 155
nm (sample r2 in table 1). A corresponding electron micrograph is
shown in figure 7a. The magnetoconductance of this ring is displayed
in figure 7b. At a temperature of 20 mK well pronounced periodic
oscillations emerge from the background with a period $\Delta B=220$
mT - 370 mT expected for the ring geometry. Also in the Fourier
spectra a clear peak is visible at approx. 3.7 T$^{-1}$ (see figure
7d, black trace). The aperiodic conductance fluctuations in the
magnetoconductance arise, as in wires, from interference effects in
the individual ring arms. These fluctuations are superimposed on the
periodic AB conductance oscillations. From the amplitude of the
oscillations we can estimate a phase coherence length of
$l_\phi\approx130$ nm by using \cite{Washburn2,Amplitude}:
\begin{equation}
\delta G=\frac{e^2}{h}\frac{L_T}{L_\phi}\mathrm{exp}(-\pi r/L_\phi)
\end{equation}
By increasing the temperature to 100 mK these periodic oscillations
disappear. This strong sensitivity on temperature is a consequence
of the temperature dependency of the phase coherence length and the
exponential damping of the oscillations. To prove the dependence of
the period on the ring diameter, i.e. the enclosed magnetic flux, we
fabricated a ring, labeled r3, with an average diameter of 190 nm.
Magnetoconductance data of this ring in a perpendicular applied
magnetic field are shown in figure 7c. At 20 mK periodic AB
oscillations are again visible but the amplitude is much smaller
than the ones observed in the smaller ring. This is expected from
the exponential suppression. Here the value of $L_\phi$ is again
$\sim130$ nm. The Fourier spectra in figure 7d (red trace) shows a
peak at approx. 6.8 T$^{-1}$ which is in very good agreement with
the expected value between 4.8 T$^{-1}$ and 9.1 T$^{-1}$. While the
lower value corresponds to the value of the inner diameter, the
higher one is extracted from the flux through the outer diameter.
Also in this ring the AB-oscillations are gone at 100 mK. A
contribution of the first harmonic of the Aharonov-Bohm effect, with
a period of h/2e, could not be observed in any of the rings
investigated. The value of the dephasing length extracted from the
Aharonov-Bohm oscillations is $\sim130$ nm for both samples (sample
r2 and r3) and thus consistent with the value extracted from the
universal conductance fluctuations.

\section{Weak localization and weak anti-localization in (Ga,Mn)As wire arrays}

The effect of weak localization \cite{Bergmann} belongs to the class
of so called time reversed interference effects. Scattered partial
waves of particles on time-reversed closed paths interfere
constructively, causing an enhanced probability of backscattering
which decreases the conductance. A perpendicular magnetic field
destroys the constructive interference and hence the quantum
mechanical correction to the conductivity and the resistance returns
within a characteristic magnetic field scale towards the Drude
value. The resulting negative magnetoresistance is the
characteristic hallmark of weak localization. As the maximum area of
the closed loops which contribute to weak localization is limited by
the phase coherence length, fits of the WL-magnetoresistance provide
another means to extract the phase coherence length. In the presence
of spin-orbit interaction, the spin part of the wave function needs
also to be taken into account.
\begin{figure}
\begin{indented}
\item[]\includegraphics[width=\linewidth]{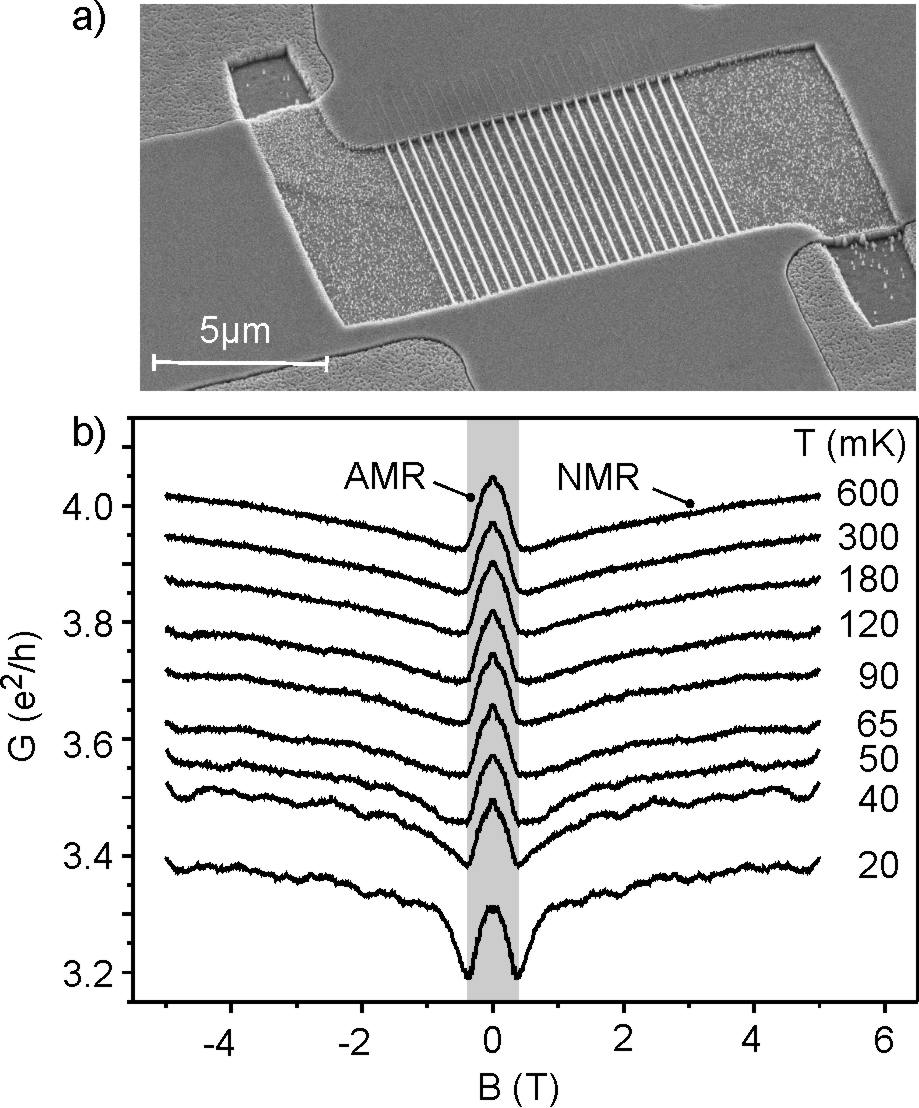}\\
\caption{a) Electron micrograph of sample a1 with 25 wires in
parallel. b) Conductance of sample a1 for different temperatures
measured in a perpendicular magnetic field. To remove the
Hall-conductance in this sample, the antisymmetric part of the
conductance was subtracted \cite{Symm}. The magnetic field range
where the magnetization is rotated from in-plane to out-of-plane is
grey-shaded and marked by AMR (anisotropic magneto resistance). The
positive slope of conductance at higher $B$-fields, the so called
negative magnetoresistance, is marked by NMR.}
\end{indented}
\end{figure}
The two partial waves on time-reversed closed paths experience a
spin rotation in opposite direction causing (partially) destructive
interference \cite{Bergmann}. So, SO interaction leads to reduced
backscattering and reverses the sign of the WL, hence called weak
antilocalization (WAL). A typical signature of WAL is a double dip
in the magnetoconductance trace \cite{Bergmann}. As a sufficient
strong magnetic field removes time reversal symmetry, the question
arises, whether weak localization can be observed at all in a
ferromagnet. In conventional ferromagnets several experimental works
explored this problem \cite{Brands1,Brands2,Ono,Aprili}, but a
direct signature of weak localization was not found. The
ferromagnetic semiconductor (Ga,Mn)As sheds new light to this
question, as the internal magnetic field in ferromagnetic
semiconductors  is quite weak compared to conventional ferromagnets.
To search for weak localization in the ferromagnetic semiconductor
(Ga,Mn)As we fabricated arrays of wires connected in parallel (see
table 1 sample a1-a2) and measured their resistance in a
perpendicular applied magnetic field. By measuring an array of
geometrically identical wires universal conductance fluctuations get
suppressed by ensemble averaging. Also here the conductance was
obtained by inverting the resistance, $G=1/R$. In figure 8a an
electron micrograph of sample a1 containing 25 nanowires in parallel
is shown. The magnetoconductance \cite{Symm} of sample a1 is shown
in figure 8b for temperatures ranging from 600 mK down to 20 mK. We
first start with the description of the dominant features observed
in experiment. The conductance maximum around zero field is due to
the so called anisotropic magneto resistance (marked AMR and grey
shaded in figure 8b). Without applied magnetic field the
magnetization of the (Ga,Mn)As wires lies in-plane. For such
in-plane magnetization the conductance is higher than for an
out-of-plane oriented magnetization \cite{Baxter}. Hence in the
low-$B$ region the magnetization is rotated from in-plane to
out-of-plane and causes the negative magnetoconductance. This
magnetic field region is highlighted by grey-shading. At higher
fields the positive slope of $G$ (marked NMR in figure 8b) is
ascribed to an increase of magnetic order \cite{Nagaev}, or to 3D
weak localization \cite{Matsukura}. At temperatures above $\sim$65
mK the different $G(B)$ traces are shifted without a noticeable
change of their shape and the AMR height scales linearly with the
background conductance. This is shown in Fig. 9, where the different
$G(B)$ traces are normalized to their $G$(3T) value.
\begin{figure}
\begin{indented}
\item[]\includegraphics[width=\linewidth]{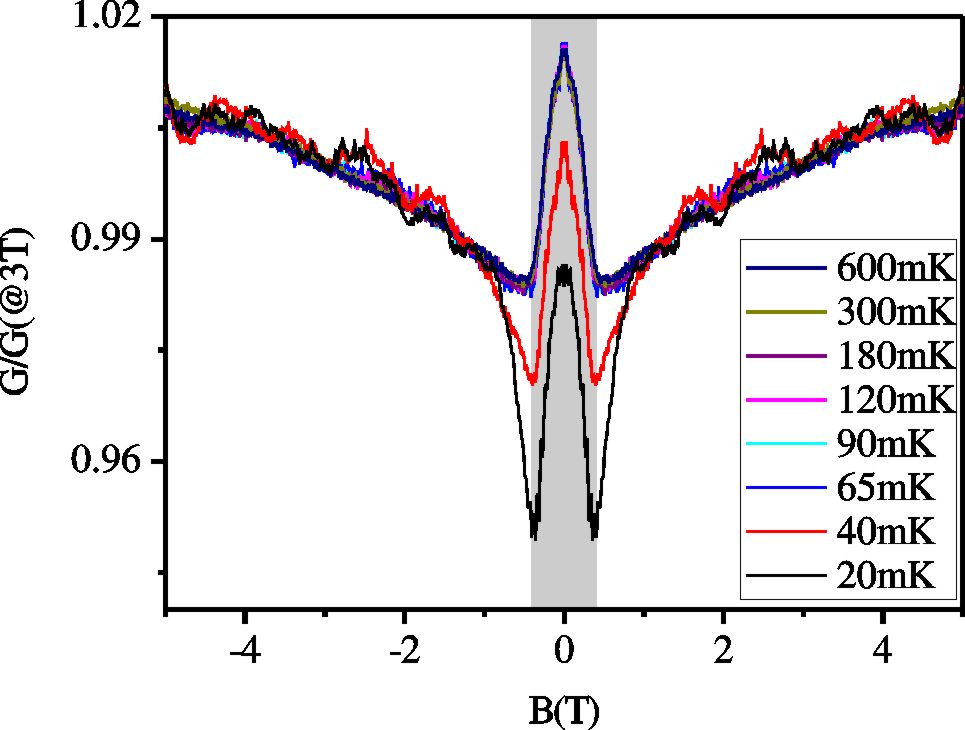}\\
\caption{Magnetoconductance of sample a1 measured at different
temperatures and normalized to the conductance value at $B$ = 3 T.
The magnetic field range where the magnetization is rotated from
in-plane to out-of-plane is grey-shaded.}
\end{indented}
\end{figure}
For temperatures between 65 mK and 600 mK all traces lie on top of
each other. This means that the conductivity at different
temperatures differs only by a constant factor $\alpha$, i.e.,
$G(T,B)=\alpha(T)G(B)$. The decreasing conductance with decreasing
temperature displayed in figure 8b is due to electron-electron
interaction (EEI). In contrast to the one-electron interference of
scattered waves which causes weak localization, EEI results from
interference of electron-electron interaction amplitudes
corresponding to successive electron-electron scattering events in
disordered systems \cite{Lee2,Dietl4}. In contrast to WL, EEI is
independent of magnetic field. In one dimensional samples the
conductance decrease due to EEI follows a $1/\sqrt{T}$ dependency
\cite{Lee2}. This dependency is shown for the three investigated
wire arrays in figure 10a. For all three wire samples the
conductivity decrease follows such a $1/\sqrt{T}$ dependency. The
effect of electron-electron interactions in 1D and 2D (Ga,Mn)As is
discussed in more detail in Ref. \cite{EEI}. Here we only note that
EEI is responsible for the shift of the traces at different
temperatures in figure 8b. At $T<65$ mK two down cusps start to
appear in the magnetoconductance of sample a1 in figure 8b at approx
$\pm$400 mT and become a prominent feature at 20 mK. This is also
seen in Fig. 9, were the different $G(B)$ are normalized to their
$G(3$T) value. While the high temperature traces lie all on top of
each other, the low temperature traces show two downward cusps with
a size becoming comparable to the AMR at 20 mK. To separate this
novel effect from the other transport contributions we have to
subtract the high temperature background containing AMR and NMR.
Here, we assume that the background scales like at temperatures
above 50 mK, i.e. that $\Delta G= G(20\mathrm{ mK})-\alpha
G(120\mathrm{ mK})$. Here we have chosen the conductance at 120 mK
as reference conductance to account for the background. The factor
$\alpha$ takes the linear scaling of AMR and NMR with the
conductance value into account. We notice that the choice of the
reference trace, we subtract, has an indiscernible influence on the
result as the high temperature traces lie all on top of each other
when normalized. The resulting traces which we ascribe to the weak
localization correction are shown in figure 10b for all 3 samples
(a1 - a2 in table 1). All three $\Delta G$ traces in figure 10b
display a local conductance maxima at $B$ = 0 and two conductance
minima at $B=\pm400$ mT. Such $\Delta G(B)$ shape is typical for the
effect of weak antilocalization in systems with spin-orbit
interaction. To extract the characteristic lengths from the data we
utilize existing theory. The correction due to WAL in a quasi 1D
system is given by \cite{Altshuler2,Pierre}:
\begin{equation}
\Delta G=g_s\frac{e^2}{h}\left[\frac{1}{2L}\left(\frac{1}{L_\phi^2}+
\frac{1}{3}\frac{w^2}{L_H^4}\right)^{-1/2}\right.-\left.\frac{3}{2L}
\left(\frac{1}{L_\phi^2}+\frac{4}{3L_{SO}^2}+\frac{1}{3}\frac{w^2}{L_H^4}
\right)^{-1/2}\right],
\end{equation}
where $g_s$ is the spin degeneracy, $L_{SO}$ is the spin orbit
scattering length describing the strength of spin orbit interaction,
and $L_H=\sqrt{\hbar/eB}$ is the magnetic length. Equation (3) is
valid as long as the quasi 1D assumption is justified:
$w,t<L_H,L_\phi<<L$.
\begin{figure}
\begin{indented}
\item[]\includegraphics[width=0.8\linewidth]{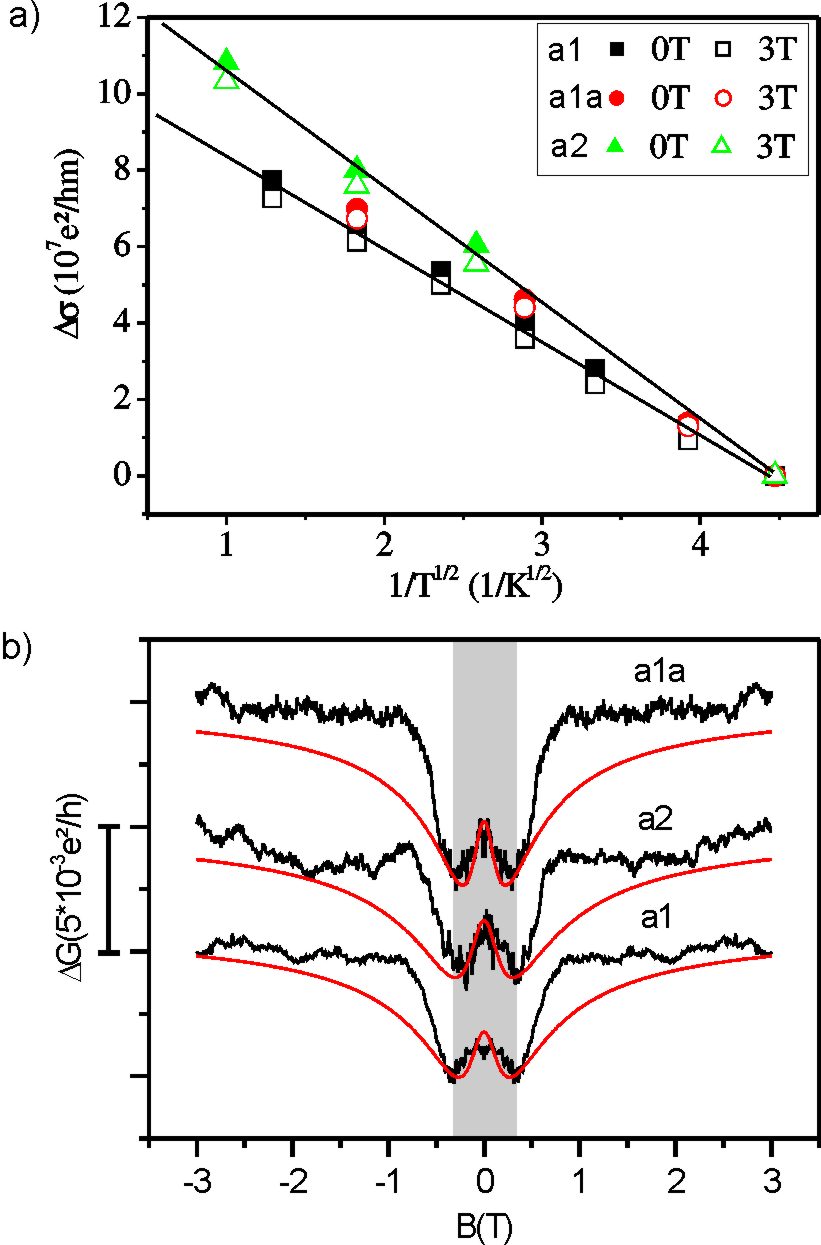}\\
\caption{a) Conductivity change of the samples a1, a1a and a2
relative to 50 mK, taken at \emph{B}=0 (solid symbols) and
\emph{B}=3 T (open symbols). The solid lines are the best linear
fits for the samples a1 and a2. b) WL contribution of the samples
a1, a1a and a2 obtained after subtracting  the 120 mK trace as
background conductance. To compare the different samples the total
$\Delta G$ was divided by the number of parallel wires. Again the
grey shaded \emph{B}-range corresponds to the regime where the
samples's magnetization follows the external field and changes
direction. The red lines are best fits to equation (3), discussed in
the text. The fit parameters were $L_{\phi}$ = 190 nm and $L_{SO}$ =
113 nm for sample a1a, $L_{\phi}$ = 160 nm and $L_{SO}$ = 93 nm for
sample a2 and $L_{\phi}$ = 150 nm and $L_{SO}$ = 93 nm for sample
a1.}
\end{indented}
\end{figure}
With the $L_\phi$ value extracted from previous experiments,
discussed above, this is justified for $T=20$ mK and $|B|<400$ mT.
As the valence band in ferromagnetic (Ga,Mn)As is spin split, the
holes are highly (but not fully) spin polarized \cite{Braden}. To
account for spin polarization, we approximated $g_s$ either by 1
(fully spin polarized) or by 2 (spin degenerate) as adjustable
parameter. As fits with $g_s=2$ can't describe the experimental
results we resort to $g_s=1$ below (The fit with $g_s=2$ for sample
a1 is shown in Ref. \cite{WL}). Corresponding fits are presented in
figure 10b as red lines. The fit parameters are $L_\phi=150$ nm and
$L_{so}=93$ nm for sample a1, $L_\phi=190$ nm and $L_{so}=113$ nm
for sample a1a, and $L_\phi=160$ nm and $L_{so}=93$ nm for sample
a2. While the fits are in good agreement with experiment for
$|B|<400$ mT they are less satisfying at higher fields. The WL or
WAL correction is, as a function of increasing $B$, more abruptly
suppressed than expected from theory. There is a striking
correlation with the magnetic field dependence of the AMR effect.
The magnetic field region where the AMR occurs is highlighted by
gray shading in figures 8b, 9 and 10b. Within this $B$-field range,
the magnetization is rotated from in-plane to out-of-plane. Once the
magnetization is out-of-plane, the WL correction drops quickly. At
the same magnetic field, the magnetic length matches wire width and
thickness, $L_H\sim w,t$. Hence, the discrepancy between fit and
experiment might be associated with dimensional crossover (1D to
3D), if $|B|$ exceeds 400 mT and equation (3) might be inapplicable.
We further note that neither the field dependent change of the
magnetization direction nor the 3/2-spin of the involved hole states
was taken into account as the theory was developed for spin 1/2
electrons in disordered metals. Especially, the latter could add a
number of additional interference diagrams not yet treated
theoretically.

While weak antilocalization has already been observed in nonmagnetic
p-type (Al,Ga)As/GaAs quantum wells \cite{Pedersen}, the observation
in a ferromagnet came as a surprise. A recent theory suggests that
the process leading to weak antilocalization is totally suppressed
in a ferromagnet due to the spin-splitting which excludes a
contribution of the so called singlet Cooperon diagram, responsible
for WAL \cite{Dugaev}. This theory was calculated for a quite strong
ferromagnet with relatively high mean free path. This is not the
case for (Ga,Mn)As and the observation of WAL in (Ga,Mn)As is not
excluded by this theory \cite{Dugaev2,Sil}.

Also in this experiment we measured the magnetoconductance of one
sample before and after annealing (Sample a1 and a1a) and compared
the resulting phase coherence length. Here, the change of the phase
coherence length was 27 \%. Again, this change can again be ascribed
to a change of the diffusion constant, while the relevant dephasing
time stayed nearly unchanged (see table 1). This underlines that the
Mn interstitials do not cause dephasing.

With the knowledge of $L_{SO}$ we now can give a more precise value
of the dephasing length $L_\phi$ extracted from the UCF measurement
discussed in section 3. The dephasing length is dependent on the
fluctuation amplitude $\delta G_{rms}$, the wire length $L$ and a
prefactor $C$ by $\delta
G_{rms}=C\frac{e^2}{h}\left(\frac{L_{\phi}}{L}\right)^{3/2}$. To
determine the prefactor $C$ knowledge of the ratio of
$L_\phi/L_{SO}$ is necessary \cite{Chandrasekhar}. For a ratio
$L_\phi/L_{SO}\approx 1.5$ (taking a spin orbit scattering length of
$\sim$100 nm and a dephasing length of $\sim$150 nm) we get a
prefactor of $C=0.58$ \cite{Chandrasekhar}. As we also deal with
relatively high magnetic field ($B=2$ T...15 T) where no weak
localization can be observed the prefactor is reduced by a factor of
$\sqrt{2}$. This factor is due to the absence of the cooperon term
at high $B$ \cite{Lee}. Taking this into account we obtain a
prefactor of $C\approx0.41$. At 20 mK this leads to dephasing
lengths for the samples w1, w2, w3, w4, w5, r1 and r1a ranging from
90 nm to 160 nm (see table 1). These values are in very good
agreement with the ones extracted from the amplitude of the
Aharonov-Bohm oscillations ($L_\phi\approx130$ nm) and the ones
obtained by fitting the weak localization correction ($L_\phi=150$
nm to 190 nm).

\section{Conclusion}

We investigated phase coherent transport in the ferromagnetic
semiconductor (Ga,Mn)As by measuring universal conductance
fluctuations, Aharonov-Bohm oscillations and weak localization. All
three methods reveal a phase breaking length of 90-190 nm at 20 mK.
By investigating universal conductance fluctuations at different
temperatures we found a temperature dependency of
$L_\phi\propto1/T^{1/2}$. As main source of dephasing
electron-electron scattering or in our case hole-hole scattering was
identified by investigating the temperature and excitation current
dependency of the UCFs. This is consistent with the quite weak
temperature dependency of $L_\phi$. The Mn interstitials do not
cause dephasing; low temperature annealing increases the dephasing
length only due to an increase of the diffusion constant, while the
corresponding dephasing time stays unchanged. The magnetotransport
in arrays of wires is modified by weak localization at temperatures
below 50 mK, showing that time reversed interference effects can be
observed in a ferromagnet. The existence of weak antilocalization
shows that spin-orbit interaction is affecting the transport in
(Ga,Mn)As. The corresponding spin orbit scattering length is always
shorter than $L_\phi$ and of order $\sim$100 nm.

\ack{We thank C. Strunk, K. Richter, V. K. Dugaev, T. Dietl and J.
Fabian for valuable discussion and the Deutsche
Forschungsgemeinschaft (DFG) for the financial support via SFB 689.}

\section*{References}



\begin{thebibliography}{00}
\bibitem{Amplitude} The conductance shown in figure 7b and 7c is the
conductance of the ring and the leads up to the voltage probe.
Equation (2) describes only the conductance oscillation of the ring
only. This issue has to be taken into account when calculating the
phase coherence length.
\bibitem{Symm} In sample a1 and a1a  the two voltage leads were
fabricated on opposite sites. So, also the Hall conductance was
measured. To remove the Hall conductance from the data only the
symmetric part of the trace was taken. As this leads to the same
result than seen on sample a2, where the voltage leads are on the
same side, physics is not changed by this procedure.
\end{thebibliography}
\end{document}